\documentclass[aps,prd,nofootinbib,superscriptaddress,twocolumn]{revtex4-1}
\pdfoutput=1

\usepackage{epsfig}
\usepackage{color}

\newcommand{\beq}{\begin{eqnarray}}
\newcommand{\eeq}{\end{eqnarray}}
\newcommand{\gsim}{\lower.7ex\hbox{$\;\stackrel{\textstyle>}{\sim}\;$}}
\newcommand{\lsim}{\lower.7ex\hbox{$\;\stackrel{\textstyle<}{\sim}\;$}}

\newcommand{\nicpb}{Laboratory of High Energy and Computational Physics, NICPB, R\"avala pst. 10, 10143 Tallinn, Estonia.}

\newcommand{\roma}{Dipartimento di Scienze Matematiche e Informatiche, Scienze Fisiche e Scienze della Terra,  Universita degli Studi di Messina, Via Ferdinando Stagno d'Alcontres 31, I-98166 Messina, Italy.}  

\newcommand{\capfe}{CAFPE and Departamento de F\'isica Te\'orica y del Cosmos, Universidad de Granada, E-18071 Granada, Spain.}

\begin{document}

\title{The 2HD+a model for a combined explanation of the possible excesses in \\ the CDF $\mathbf{M_W}$ measurement  and $\mathbf{(g-2)_\mu}$  with Dark Matter}

\author{Giorgio Arcadi}
\affiliation{\roma}

\author{Abdelhak Djouadi}
\affiliation{\capfe}
\affiliation{\nicpb}

\begin{abstract}

The new measurement of the $W$ boson mass performed by the CDF experiment at the Tevatron shows a significant deviation not only with the expectation in the Standard Model but also with other precision measurements performed at LEP, the Tevatron and the LHC. We nevertheless take this new measurement at face value and interpret it as an effect of new physics. We particularly try to link it with other possible anomalies such as the recent muon $g-2$ and consider a scenario that addresses some shortcomings of the Standard Model.  We show that a model with two doublets and a light pseudoscalar Higgs fields, supplemented by a stable isosinglet fermion, can simultaneously explain the possible $M_W$ and $(g-2)_\mu$ anomalies and accounts for the weakly interacting massive particle that could be responsible of the dark matter in the universe. 

\end{abstract}

\maketitle

%%%%%%%%%%%%%%%%%%%%%%%%%%%%%%%%%%%%%%%%%%%%%%%%%%%%%%%%%%%%%%%%%%%%%
\section{Introduction}

\noindent The CDF experiment at the Tevatron has released a new measurement of the $W$ boson mass \cite{CDF} 
\beq 
M_W = 80.4335 \pm 0.0094~{\rm GeV}\,. 
\eeq
On the one hand, the combined statistical and systematical errors on this new measurement is  smaller than that of the current world average value obtained when combining all former measurements from LEP, Tevatron and the LHC,  $M_W = 80.379 \pm 0.012 ~{\rm GeV}$ \cite{PDG}. The  central value of this average is more than 50 MeV lower than the CDF new value and,  when one combines all available data, one obtains  $M_W = 80.4133 \pm 0.0080~{\rm GeV}$ \cite{Jorge-new}.
 
On the other hand, the new CDF value deviates from the expectation in the Standard Model (SM), as a recent global fit of all electroweak precision data gives \cite{Jorge-old} 
\beq
M_W = 80.3545 \pm 0.0057~{\rm GeV}\, , 
\eeq
and this deviation from the theoretical prediction is huge, slightly more than $7 \sigma$. Even if one compares the prediction with the new averaged $M_W$ value, the deviation is still at a very high level \cite{Jorge-new}.  This new and unexpected development calls for great caution and confirmation  (as it is customary to say, extraordinary claims require an extraordinary evidence) and, at least, a careful understanding of the differences between the various measurements is mandatory before any firm conclusion is made. 

Nevertheless, as the main mission of a particle theorist is to interpret the experimental data without any qualms, one should take this new result at face value, put it in perspective and interpret it in the context of physics beyond the SM and/or infer its possible implications, as it was already done in many very recent analyses \cite{papersMW1,papersMW2}. In particular, one should at least try to relate it to other observed anomalies and embed it in model extensions that address important shortcomings of the SM. 

It would be particularly welcome if the new $M_W$ value is connected with another recent discrepancy also observed at Fermilab, the one affecting the muon anomalous magnetic moment released a year ago by the Muon g--2 collaboration \cite{g-2} and which exhibits a $4.2\sigma$ deviation from the SM expectation.  There are also standing anomalies, albeit weaker, occurring in $B$-meson obser\-vables and some of them are also associated with muons, e.g.~the semi-leptonic $b\to s \mu^+ \mu^-$ decay rate \cite{Bmumu}. 

In fact, in a recent paper \cite{paper0}, we correlated the above two anomalies in the context of a new physics model that also addresses a main concern of the SM, namely its inability to account for the dark matter (DM) in the universe. The model is based on an extension of the SM Higgs sector to contain two Higgs doublet fields and a light pseudoscalar $a$ state with enhanced couplings to muons \cite{2HDa,PhysRep,2HDa2}.  This particle then minimally couples to an additional SU(2) isosinglet fermion which is assumed to be stable and forms the DM. We have shown that such a 2HD+a model can easily cope with all existing constraints from collider and astroparticle physics and, at the same time, explains the deviations observed in the measurement of the $(g -2)_\mu$ and BR($b \to s \mu^+ \mu^-)$. 

In this brief note, we reconsider this 2HD+$a$ model and relax an assumption made to ease the numerical analysis, namely that the heavier Higgs states, the CP-even $H$, CP-odd $A$ and two charged $H^\pm$ bosons, are degenerate in mass to comply with electroweak precision data \cite{degenerate}. This will not affect the aspects related to DM and flavor physics, but introduces a correction to the $\rho$ parameter that modifies the $W$ mass value. We will show that the parameters of the model can be chosen in such a way that the CDF measurement is recovered without significantly impacting  the other observables including the $(g -2)_\mu$ excess  and allowing for a good DM candidate.   

In the next section, we summarize our model and present two benchmarks in which the $(g\!-\!2)_\mu$ excess is resolved with all collider and astroparticle physics constraints satisfied. In section 3, we discuss new contributions to $M_W$ and show that the CDF value is reproduced in these benchmarks. A  conclusion is given in section 4. 

%%%%%%%%%%%%%%%%%%%%%%%%%%%%%%%%%%%%%%%%%%%%%%%%%%%%%%%%%%%%%%%%%%%%%
\section{The 2HD+a model and Dark Matter}

Models with two-Higgs doublet (2HDM) fields $H_1$ and $H_2$, acquiring non-zero expectation values $v_1$ and $v_2$ with $\sqrt{v_1^2+v_2^2}=v\simeq 246~{\rm GeV}$ and a ratio denoted by $\tan\beta =v_1/v_2$, are interesting and widely discussed extensions of the SM \cite{2HDM}. They lead to a richer Higgs spectrum consisting of two CP--even $h,H$ bosons, with $h$ assumed to be the state with a mass of 125 GeV observed at the LHC, a CP-odd  $A$ and two charged $H^{\pm}$ bosons. The main motivation of a 2HDM plus a light pseudoscalar $a$ boson is that it allows, in addition, to induce a gauge invariant interaction between the gauge singlet $a$ boson and pairs of SM fermions, as well as pairs of the fermionic singlet DM $\chi$, through the mixing with the $A$ state, described by the angle  $\theta$ defined by 
\beq
\tan2\theta= 2 \kappa v/(M_{A}^2-M_{a}^2), 
\eeq
with $\kappa$ the coefficient of the term coupling the two doublets $H_1,H_2$  with the original singlet field. For $M_A \! \gg \! M_a$ and for a strong mixing $\sin2\theta \! \approx \!1$, there is an upper bound on $M_A$ from the requirement of perturbative unitarity in Higgs scattering amplitudes, $M_A \! \lsim \! 1.4$ TeV \cite{PhysRep}. 

In the simplest case in which the 2HDM sector  is CP-conserving and possesses a $Z_2$ symmetry that forbids tree--level flavor-changing neutral currents, the model is characterized by the following set of input parameters: the five Higgs masses  $M_h$, $M_H$, $M_{H^{\pm}}$, $M_A$ and $M_a$, the mixing angles $\theta$ among the $a,A$ and $\alpha$ among the CP-even $h,H$ states, $\tan\beta$ and three parameters of the scalar potential which enter only in the self-couplings among the Higgs bosons. The only requirement we will make on these last parameters is that they should lead to a very small coupling among the $haa$ states, $\lambda_{haa} \lsim 10^{-3}$. In order to cope with the LHC constraints which force the observed $h$ particle to have almost SM-like couplings to fermions and weak bosons \cite{LHC-h}, one  can impose the so-called alignment limit, $\alpha=\beta-\frac12 \pi$. 

Compared to the SM case \cite{TomeI}, the Higgs couplings to the SM fermions are proportional to a factor $g_{\Phi ff}$ where in the alignment limit one has $g_{hff}\!=\!1$ for the light $h$ while the ones 
of the other Higgs states are given by 
\begin{equation}
    g_{Hff}=\xi_f, \quad g_{Aff}=\cos\theta \; \xi_f, \quad g_{aff}=-\sin\theta \; \xi_f \; , 
\end{equation}
with the coefficients $|\xi_f|\! = \! \tan\beta$ or $\cot\beta$ depending on the type of the considered 2HDM \cite{2HDM}. In the present situation, as we need enhanced couplings to the isospin $-\frac12$  muons, we will discuss only the so-called Type--II scenario with $|\xi_\tau| \! = \!|\xi_b| \! = \! 1/|\xi_t| \! = \tan\beta$ and the lepton--specific or Type--X scenario with $ |\xi_\tau| \! = \! 1/|\xi_b| \! = \! 1/|\xi_t|= \tan\beta$. In both cases the non-standard Higgs states will have strongly enhanced couplings to charged leptons for values $\tan\beta \! \gg \!1$ and, in Type-II, also to bottom quarks. 

Finally, for what concerns the DM aspect, we will assume the presence of a stable fermion $\chi$ which is a SM singlet and hence does not couple to gauge bosons and couples to Higgs bosons only in pairs. In particular, there are no $\chi$ couplings to the CP-even $h,H$ bosons while its couplings to the two pseudoscalar bosons is given by 
\begin{equation}
\mathcal{L}_{\rm DM}=g_\chi \left(\cos\theta a+\sin\theta A\right) \bar \chi i \gamma_5 \chi \, . 
\end{equation}

This model, recently discussed in Ref.~\cite{paper0} to which we refer for further details,  has remarkable virtues that we briefly summarize in the following.   

-- It passes all collider constraints on the 2HDM bosons if one is in the alignment limit which forces the $h$ particle to be SM-like \cite{LHC-h} and the heavier $H,A$ and $H^\pm$ states to be heavy enough to escape direct detection at the LHC \cite{LHC-HA,LHC-H+}.  In the case of the Type-II model, assuming $M_A \! \approx \! M_H \approx \! M_{H^\pm}$ which is the simplest way to avoid the occurrence of large corrections to electroweak observables, one can adapt  to the present 2HD+$a$ case the constraints on the $[M_A,\tan\beta]$ plane derived for the MSSM \cite{MSSM}. In the MSSM, values $\tan\beta \gsim 10$ are excluded for 
$M_H\!=\!M_A \lsim 1$ TeV from $pp \! \to \! H/A \! \to \! \tau^+\tau^-$ searches \cite{LHC-HA}, while values $\tan\beta \gsim 50$ are excluded by $pp \to H^\pm \to tb$ searches for $M_{H^\pm} \lsim 700$ GeV \cite{LHC-H+}. However, in our case, given the presence of additional decay channels, such as $H\! \rightarrow \! aa$, $A\! \rightarrow \! ah$, $H \! \rightarrow \!  Za$ and $H^\pm \to aW$, we expect weaker limit than in the MSSM. In Ref.~\cite{paper0}, we have thus assumed the value $M \gsim 1$ TeV.  In the Type--X  model, since all production vertices of the new Higgs bosons at the LHC are suppressed at high $\tan\beta$,  the interpretation of the $g-2$ excess is much easier and it is possible to lower the value of $M$ down to a few hundreds of GeV despite of the constraints from $H/A$ production.

For the light $a$ state, one needs to have small $abb$ and $a\tau\tau$ couplings to evade LEP bounds \cite{LEPa} and a tiny $haa$ coupling, $\lsim 10^{-3}$, to sufficiently suppress the LHC constrained additional $h \to aa $ boson decays \cite{hinvisible}.   In the range of small Higgs boson masses considered for the Type--X model, severe constraints come also from lepton universality in $Z$-boson and $\tau$-lepton decays \cite{PDG}.

Flavor physics is also a relevant source of constraints as the rate of decays of $K$ mesons and above all,  $B$ mesons \cite{Bphys} are sensitive to the presence of additional Higgs bosons. In our scenario, the $B_s \to \mu^+\mu^-$ process is particularly important since it could receive significant contributions from the light $a$ state, possibly emitted on mass-shell, with enhanced couplings to muons \cite{BmumuTH}.   Particularly relevant is also, for the Type--II case, the bound  $M_{H^\pm}> 570\,\mbox{GeV}$~\cite{Misiak:2017bgg} from the $b\to s\gamma$ radiative decay.

For a few GeV mass  and in Type-II and Type-X or scenarios with large enough $\tan\beta$ values, the $a$ boson can give the required contribution to the $(g-2)_\mu$ through  Barr--Zee type diagrams occurring at the two--loop level \cite{Barr-Zee} that allow to explain the $4.2\sigma$ deviation of the Fermilab measurement from the SM expectation. There are also one loop contributions of the $a$ and 2HDM Higgs states but they are suppressed compared to the above. 

\begin{figure*}[t]
    \centering
\mbox{    \includegraphics[width=0.43 \textwidth]{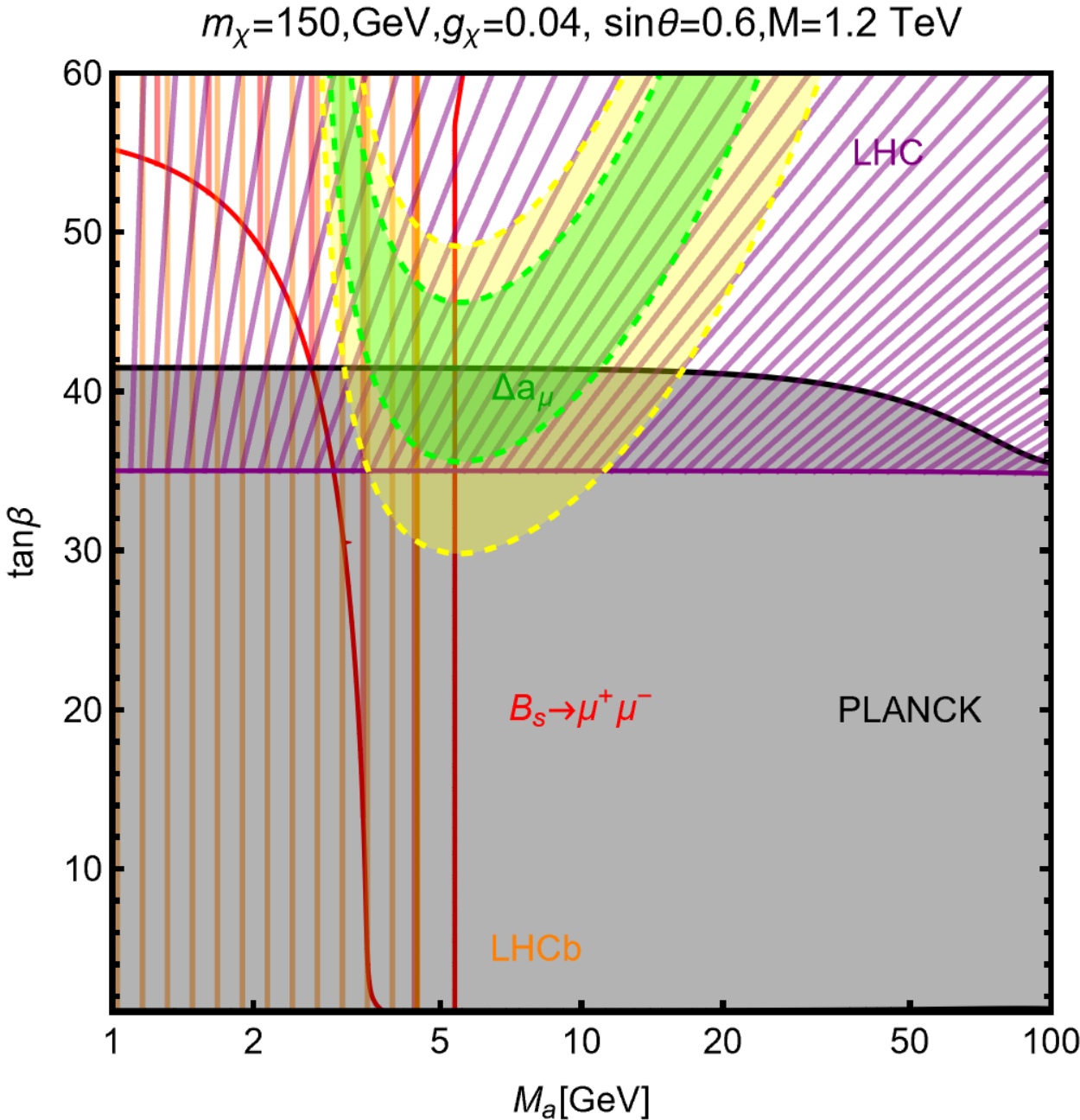}~~~
    \includegraphics[width=0.43\textwidth]{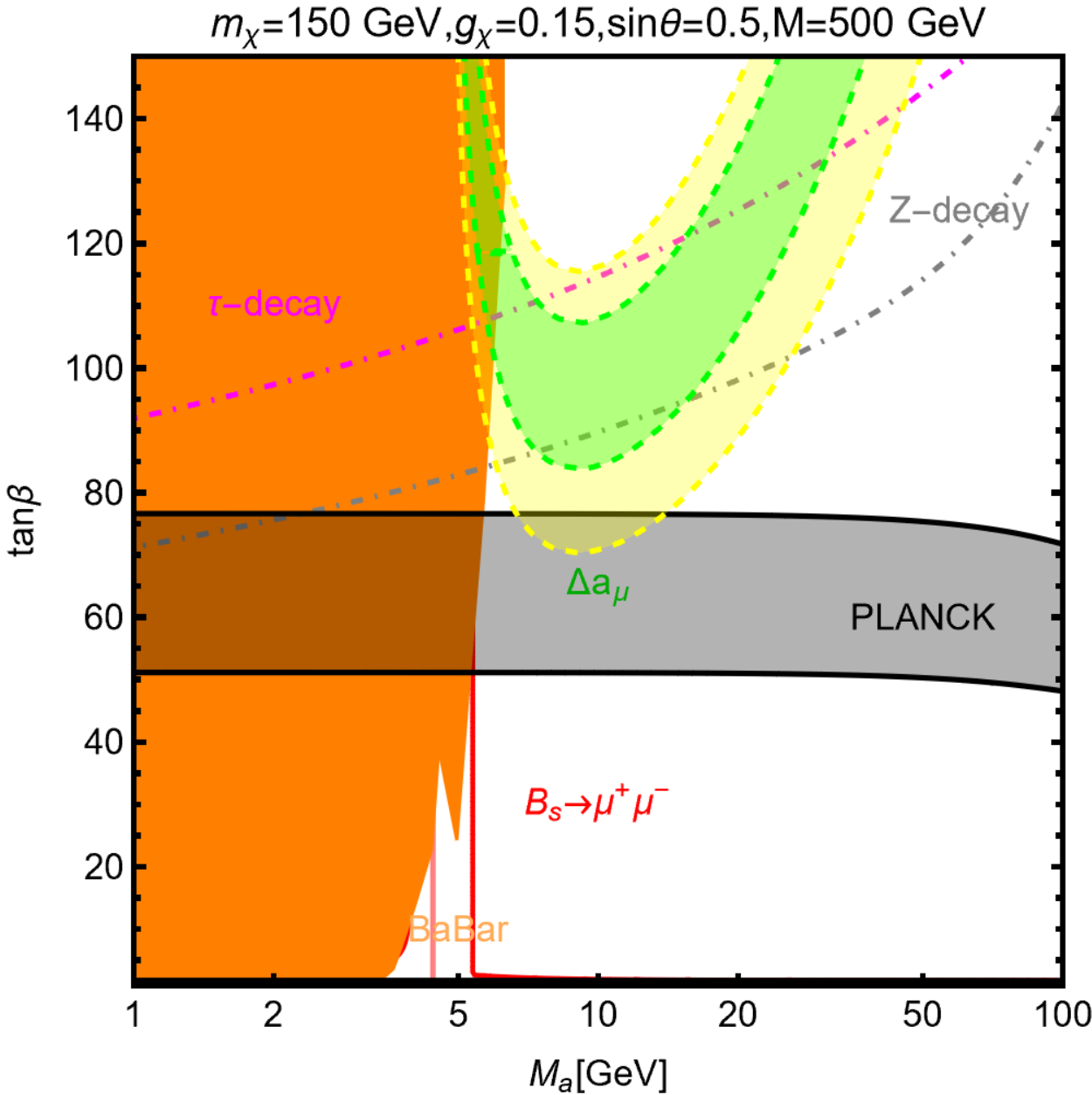}
    }
    \caption{A summary of the various constraints in the $[M_a,\tan\beta]$ plane for the 2HD+$a$ model in the Type-II  (left panel) and Type-X (right panel) scenarios for the assignments of the $(m_\chi,g_\chi,\sin\theta)$ parameters displayed on top of the corresponding panels. The colored black bands correspond to regions in which the correct DM relic density is obtained, the green and yellow bands are when a fit of the $(g-2)_\mu$ measurement is achieved within $1\;\sigma$ and $2\;\sigma$, the red areas are those in which the decay rate of $B_s \! \rightarrow \! \mu^+ \mu^-$ exceeds the experimental limit, the purple hatched region is excluded mainly by the LHC searches of $H/A$ bosons decaying into $\tau$ pairs. In the bottom panel, the regions above the dot-dashed gray (magenta) contours are excluded by bounds on lepton universality violation in $Z$-boson ($\tau$) decays. We have taken the 2HDM mass scale to be $M\!=\! 1.2\; (0.5)$ TeV for the Type-II (Type-X) cases.}
    \label{fig:ptypeII}
\vspace*{-3mm}
\end{figure*}

Finally, for what concerns the DM issue, two ingredients make that the $\chi$ state with a ${\cal O}($100 GeV) mass provides the correct cosmological relic density as measured by the PLANCK satellite \cite{PLANCK} and chiefly passes the stringent constraints from experiments in direct detection such as XENON1T \cite{XENON1T} and indirect detection such as FERMI-LAT \cite{FERMI-LAT}. First, $\chi$  does not couple to the CP-even $h$ and $H$ hence forbidding spin-independent interactions at tree--level and, second, new $\chi \chi$ annihilation channels involving the light $a$ are present and allow for an efficient annihilation into lighter particles.    

Following Ref.~\cite{paper0}, we briefly recapitulate the impact of all these constraints in Fig.~1 for the Type--II (left panel) and lepton-specific (right panel) scenarios. We have fixed $\sin\theta$ and the parameters $m_\chi,g_\chi$ related to the DM particle to the values given in the top of the figures and, for the Higgs sector, assumed the alignment limit and a common mass for the heavier Higgs bosons, $M_A=M_H=M_{H^\pm}=M=1.2$ TeV in the Type--II case where the LHC bounds from direct searches of the $A/H$ and to a lesser extent $H^\pm$ are severe at high $\tan\beta$, and $M=0.5$ TeV in the lepton-specific or Type-X case, as these Higgs searches are inefficient and much lower Higgs mass values are allowed. We then vary the parameters $M_a$ and $\tan\beta$ and apply all constraints from the LHC, $(g-2)_\mu$, flavor physics as well as direct and indirect detection of the DM particle with a correct relic abundance. 

As can be seen, for the chosen values of the various inputs, there are regions of the $[M_a, \tan\beta]$ parameter space, in particular for small $M_a$ and high $\tan\beta$ values,  in which  all astroparticle and collider constraints are satisfied, while leading to a  $(g-2)_\mu$ value close to the measured one.

In the previous discussion, a major assumption has been made to ease the numerical analysis, namely that the heavier $H,A,H^{\pm}$ states are degenerate in mass, $M_H=M_{A}=M_{H^\pm}=M$,  as is naturally the case in some models, such as the MSSM in the decoupling regime \cite{MSSM}. This was the simplest way to avoid large contributions to electroweak precision observables \cite{degenerate} and, in particular, the $W$ mass to which we turn our attention now. 

%%%%%%%%%%%%%%%%%%%%%%%%%%%%%%%%%%%%%%%%%%%%%%%%%%%%%%%%%%%%%%%%%%%%%%%
\section{Impact on the W boson mass}

The leading radiative corrections to $M_W$ can be approximated by the one affecting the $\rho$ parameter which measures the strength of the neutral to the charged currents ratio at zero--momentum transfer \cite{Veltman}, 
\beq 
\frac{\Delta M_W}{M_W} \simeq \frac12 \frac{M_W^2}{2M_W^2-M_Z^2} \Delta \rho \approx \frac34 \Delta\rho \, , \nonumber \\
\Delta \rho = \frac{\Pi_{WW}(0)}{M_W^2}  - \frac{\Pi_{ZZ}(0)} {M_Z^2} \, .~~~~~~
\eeq
where $\Pi_{VV}$ are the transverse parts of the $V\!=\!W,Z$ boson self-energies. In our 2HD+$a$ model, the additional contributions due to the extra Higgs states (we ignore the SM-like contribution of the $h$ boson which is included in the fit of the SM data) is given, in the alignment limit,  by (see also Ref.~\cite{Drho} in which the contributions in a 2HDM  were first discussed and Ref.~\cite{papersMW1} in which they were interpreted in this new $M_W$ context):
\beq
\Delta \rho = \frac{\alpha}{16 \pi^2 M_W^2 (1 -M_W^2/M_Z^2)} \big[] 
f(M^2_{H\pm},M^2_H)\nonumber \\
+ \cos^2\theta f(M^2_{H\pm},M^2_A) + \sin^2\theta f(M^2_{H\pm},M^2_a) \nonumber \\
- \cos^2\theta f(M^2_A,M^2_H) - \sin^2\theta f(M^2_a,M^2_H) \big] \, , 
\eeq 
where $\alpha$ is the fine structure constant and $\theta$ the $Aa$ mixing angle. The function $f$ is given by 
\beq
f(x,y) = x+y- \frac{2 x y}{x-y} \log \frac{x}{y} \, , 
\eeq
and vanishes if the two particles running in the loop  are degenerate in mass $f(x,x)=0$ while, in the limit of a large mass splitting, one has $f(x,0)=x$ instead. Hence, in the case where the members of an SU(2) doublet have masses that are quite different, contributions which are quadratic in the mass of the heaviest particle appear. 

Let us first evaluate the impact of the new $M_W$ measurement in the Type-II benchmark. As discussed in Refs.~\cite{PhysRep, 
2HDa2}, in Type-II 2HDMs and for large Higgs masses, only a limited  mass splitting between the $H/A/H^\pm$ states is allowed when one requires compatibility with theoretical constraints from perturbativity and unitarity such as those that apply on the quartic couplings of the scalar potential. This is particularly true when the  alignment limit,
$\cos (\beta\!-\!\alpha)=0$, is imposed. 

\begin{figure}[!h]
    \centering
    \includegraphics[width=0.95\linewidth]{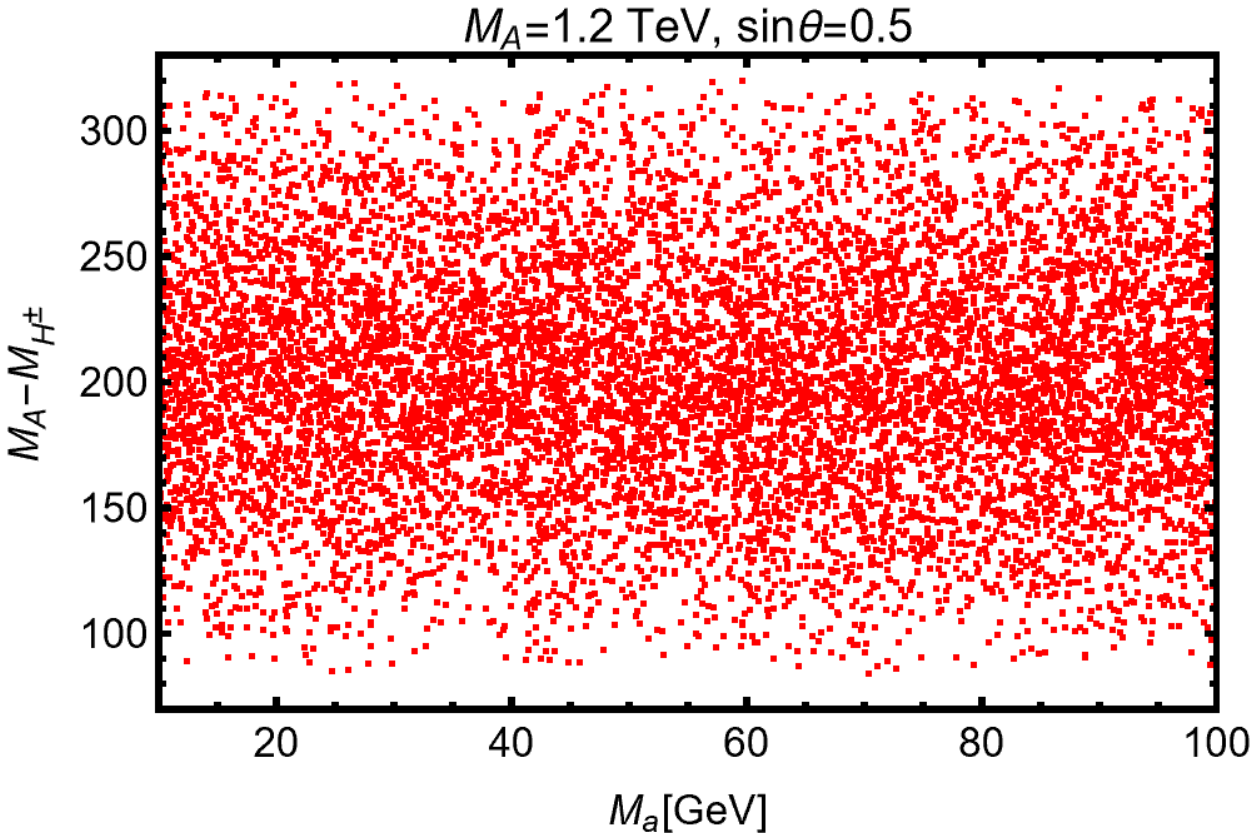}
    \vspace*{-1mm}
    \caption{Range of values of $M_{H^{\pm}}\!-\!M$ as a function of $M_a$ and considering $\tan\beta\in \left[1,60\right]$ and $M\!=\!1.2\,\mbox{TeV}$ in Type-II models that  accounts for the deviation in $M_W$ as  reported by CDF.}
    \label{fig:MW}
    \vspace*{-1mm}
\end{figure}

In the present case, we will simply consider the minimal option of a splitting between $M_A\!,\!M_H\!$ and $M_{H^{\pm}}$, assume again the alignment limit and perform a scan over the possible values of $M_{H},M_{H^{\pm}}$ and , also varying $M_a$ in the range of a few 10 $\mbox{GeV}$ and $\tan\beta$ in the perturbative range allowed for the bottom Yukawa coupling, 
\begin{eqnarray}
    & M_a \in \left[10,100\right]\,\mbox{GeV}\, ,  \,\,\,\tan\beta\in \left[1,60\right] \, , 
\end{eqnarray}
accounting also for the theoretical bounds on the couplings of the scalar potential \cite{2HDa2}. The result is shown in Fig.~\ref{fig:MW} for $M\!=\!M_H\!=\!M_A\!=1.2$ TeV and  $\sin\theta=0.5$.

Almost independently from the value of $M_a$, at least within the considered range, the CDF measurement can be successfully interpreted simply by considering a splitting $M\!-\!M_{H^{\pm}} \approx\! 100$--300 GeV. This Higgs mass difference has no impact on the observables shown in Fig.~\ref{fig:ptypeII}, which then remain the same also in the light of the  deviation from the SM of the CDF $M_W$ value, if indeed confirmed.

\begin{figure}
    \centering
%\mbox
{    
\includegraphics[width=0.36\textwidth]{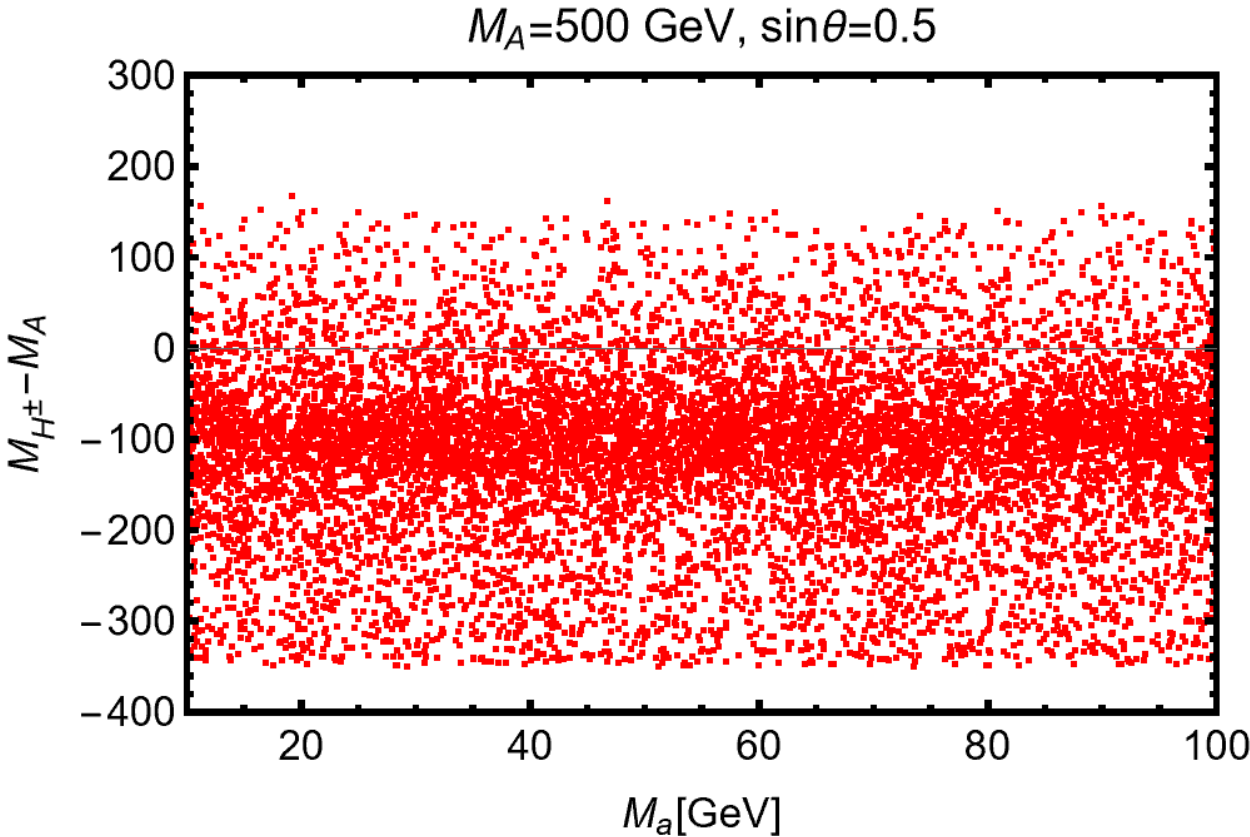}\\[1mm]
\includegraphics[width=0.36\textwidth]{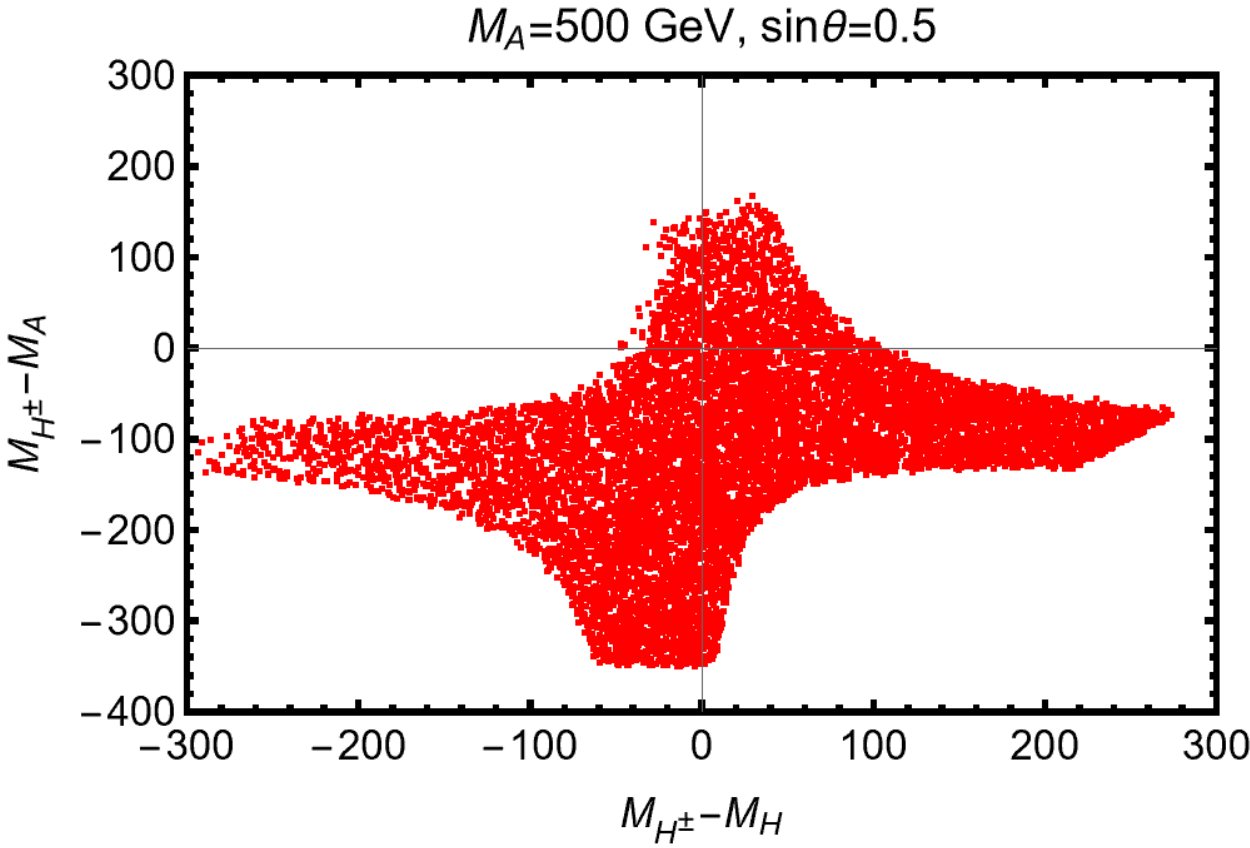}\\[1mm]
      \includegraphics[width=0.36\textwidth]{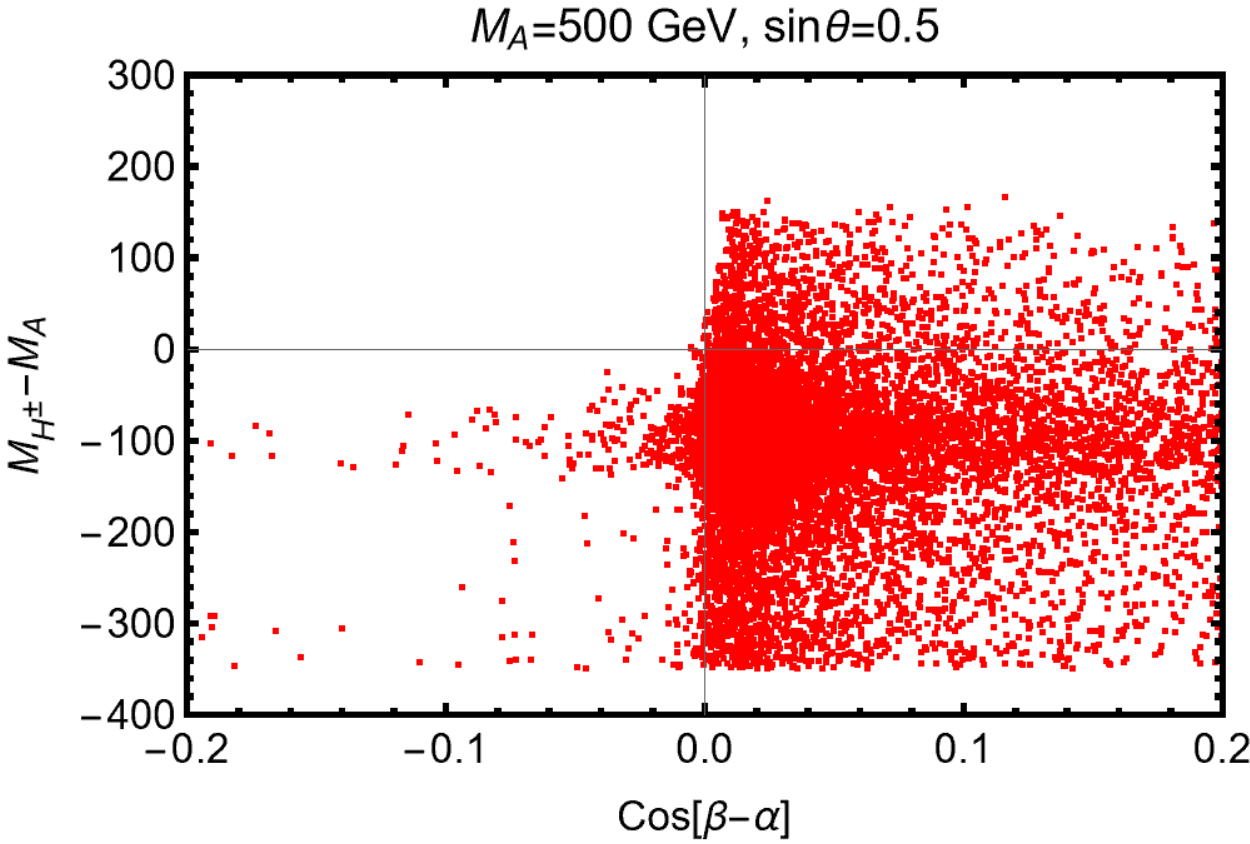} }
      \vspace*{-3mm}
    \caption{Outcome of the scans of the type-X 2HD+$a$ parameter space in the planes $[M_a, M_{H^\pm}\!-\!M_A]$ (left),  $[M_{H^\pm}\!-\!M_H, M_{H^\pm}\!-\!M_A]$ (middle) and $[\cos(\beta\!-\!\alpha), M_{H^\pm}\!-\!M_A$] (right)  when  fixing the values $M_A\!=\!500\,\mbox{GeV}$ and $\sin\theta\!=\!0.5$. The points shown are compatible with theoretical constraints on the scalar potential and give a value of $M_W$ compatible with the CDF measurement.}
    \label{fig:ptypeXscan}
\vspace*{-3mm}
\end{figure}

Moving to the Type-X scenario, a slightly more extensive analysis can be conducted. Indeed, theoretical constraints allow for higher Higgs mass splitting in this case and,  moreover, small deviations from the alignment limit can be considered. Assuming a fixed value for the mass of the heavy pseudoscalar Higgs state, $M_A=500\,\mbox{GeV}$ and choosing again $\sin\theta=0.5$,  we have varied the other Higgs sector parameters as follows
\begin{eqnarray}
    & M_a \in \left[10,100\right]\,\mbox{GeV}\, , \,\,\,\tan\beta\in \left[1,150\right]\, , \nonumber\\
    & M_{H,H^{\pm}} \in \left[100,1000\right]\,\mbox{GeV}\, , \,\,\,|\cos(\beta-\alpha)|<0.2 \, . 
\end{eqnarray}

We display in the three panels of Fig.~\ref{fig:ptypeXscan}, the model points that are compatible with the theoretical constraints discussed in Refs.~\cite{PhysRep,2HDa} and giving a deviation to $\Delta \rho$ compatible with the CDF measured $M_W$ value.

In the left panel of Fig.~\ref{fig:ptypeXscan}, we simply repeat the Type-II analysis and show the results in the $[M_a, M_{H^\pm} -M]$ plane. In the Type-X scenario, in addition to the fact that smaller 2HDM Higgs masses can be considered, the range of mass splitting that allows to explain the new $M_W$ value is relatively wider; again, the impact of $M_a$ in the considered 10--100 GeV range is rather small. 

In the middle panel, we show the simultaneous values of  $M_{H^\pm}\! -\!M_A$ and $M_{H^\pm} \!-\!M_H$ which allow to obtain the desired contribution to the $W$ boson mass. And these include small $M_{H^\pm}$ values, close to the exclusion limit from the LEP experiment, $ M_{H^\pm} \gsim M_W$ \cite{PDG}. Finally, the right panel shows that a small deviation from the alignment limit, $\cos(\beta -\alpha) \neq 0$,  can still  account for the CDF  value even if the $M_{H^\pm}\! -\!M_A$ difference is significant.

Before closing this discussion, let us note that the implications for the $M_W$ value of the mass splitting of the scalars in a 2HDM has been also discussed in the series of papers of Ref.~\cite{papersMW1} that appeared after shortly the CDF announcement.  Most of these analyses made use of the Peskin-Takeuchi formalism \cite{STU} in which, besides the $T$ parameter that is equivalent to the $\Delta \rho$ correction,  $T \propto \Delta \rho$, and which has the biggest impact, also the $S$ parameter was considered (the parameter $U$ gives too small contributions that were neglected in most cases).  

While we qualitatively agree with the results of Refs.~\cite{papersMW1}, the situation is more complicated in our case since we have the additional contribution of the light $a$ state which, from the start, has a very large mass  splitting compared to $M_A, M_H$ and $M_{H^\pm}$. The presence of this light $a$ boson will require a slightly smaller mass splitting $M_H\!-\!M_{H^\pm}$ and $M_A-M_{H^\pm}$ to comply with the new $M_W$ value, compared to the 2HDM case.

\begin{figure}[!h]
    \centering
    \includegraphics[width=0.7\linewidth]{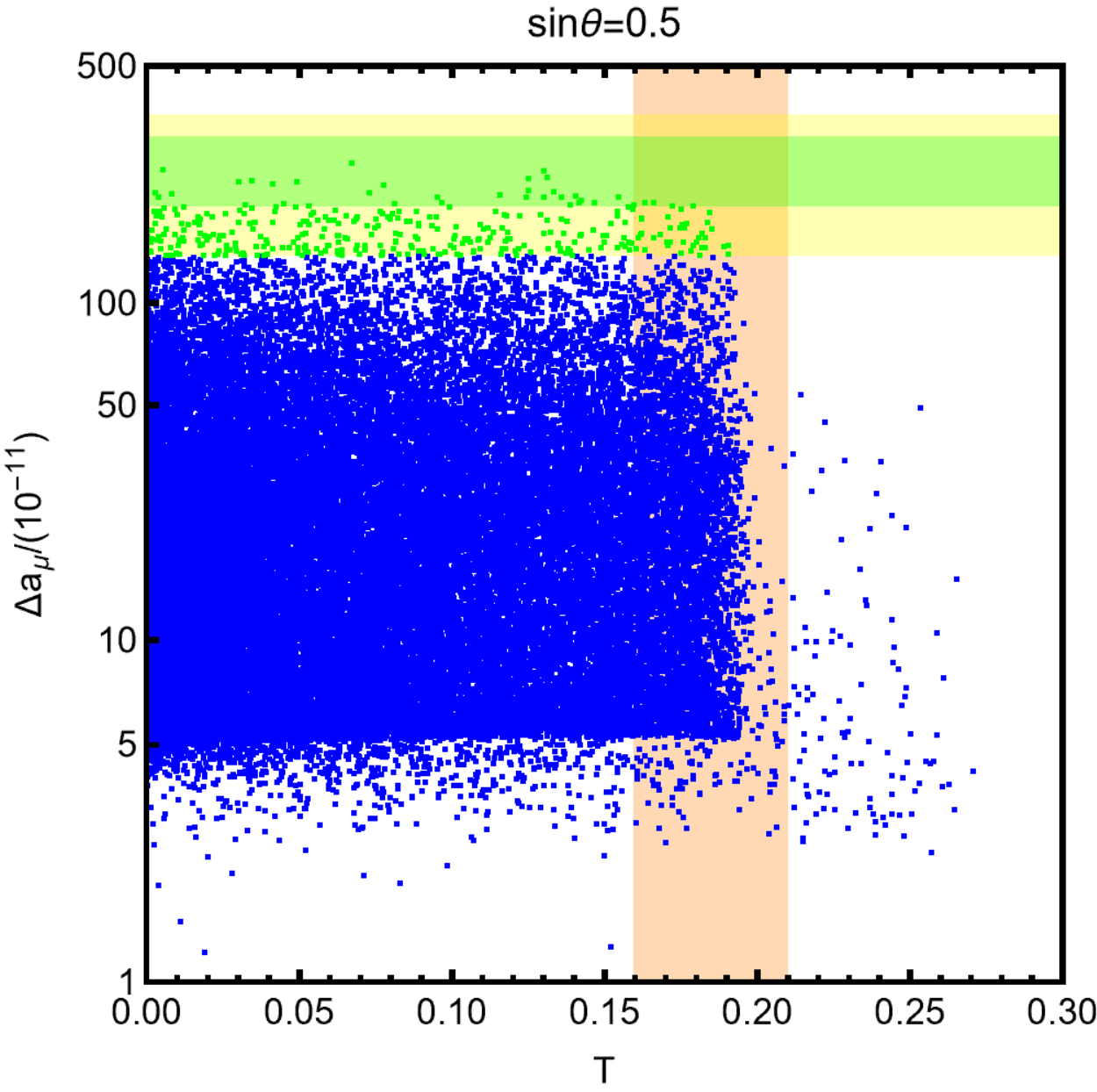}
\vspace*{-3mm}
    \caption{Type-X model points in the   $[T,\Delta a_\mu/(10^{-11}]$ bidimensional plane that comply with the various constraints. The green and yellow horizontal bands represent the $1\sigma$ and $2\sigma$ allowed  ranges for $\Delta a_\mu$ while the orange vertical band represent the range of values of $T$ given in eq.~(\ref{eq:Trange}).}
    \label{fig:CDFg2}
\vspace*{-1mm}
\end{figure}

Finally, to conclude our analysis with an explicit illustration, we show in Fig.~\ref{fig:CDFg2} the correlation between the deviations of the CDF $M_W$ and the muon $(g-2)$ measurements from the SM expectation 
using the $S,T,U$ formalism. We display the same model points already presented in Fig.~\ref{fig:ptypeXscan} but in the $\left[T,\Delta a_\mu/(10^{-11} \right]$ bidimensional plane, with $a_\mu =\frac12 (g-2)_\mu$. The figure highlights the $1\sigma$ (in green) and $2\sigma$ (in yellow) horizontal regions corresponding to the $(g-2)_\mu$ anomaly and, as a vertical band depicted in orange, the range of values of the $T$ parameter 
\beq
T \in [0.16, 0.21] \, , 
\label{eq:Trange}
\eeq
which has been proposed in Ref.~\cite{Babu:2022pdn} to explain the CDF measurement of $M_W$.  As can be seen, it is in general rather difficult to simultaneously accommodate the CDF $M_W$ and the Fermilab  $(g-2)_\mu$ measured values, but a non zero overlapping region nevertheless exists.
                    
\section{Conclusions} 

The new measurement of the $W$ mass reported by the CDF collaboration features a large deviation from the theoretical expectation in the  SM, but is also widely different from  previous measurements made at LEP, the Tevatron and LHC. While a detailed and careful analysis of the various systematical errors that affect the different $M_W$ measurements is required (and a measurement from the CMS collaboration would be welcome) before drawing a definite conclusion, one can nevertheless  speculate about the possibility that this deviation could be due to new physics beyond the SM and, eventually, relate it to additional anomalies observed in other measurements. 

In this brief note, we have considered a two Higgs  doublet model supplemented by a light pseudoscalar Higgs boson to which we add a new stable singlet fermion to  account for the dark matter. We show that under some conditions, the anomalous contribution to $M_W$ can be reproduced by allowing some  mass non-degeneracy for the heavier Higgs bosons.  At the same time, one can explain the excess observed in the value of the muon anomalous magnetic moment (if it is indeed also real) and comply with $B$-meson physics constraints such as those originating from the $b \to s\gamma$ and $b \to s \mu^+ \mu^-$ decay rates.  All this while satisfying the direct and indirect light and heavy Higgs searches at the LHC and elsewhere, as well as the astrophysical constraints from direct and indirect detection of a DM with the observed cosmological density.  

Such a scenario can be probed already at the next LHC run and with a slight increase of sensitivity of astrophysical experiments searching for thermal dark matter.\smallskip   

\noindent {\bf Acknowledgements:} A discussion with Jorge de Blas on Ref.~\cite{Jorge-new} is acknowledged. AD is supported by the Estonian Research Council (ERC) grant MOBTT86 and by the Junta de Andalucia through the Talentia Senior program and the A-FQM-211-UGR18 and P18-FR-4314 grants.

%%%%%%%%%%%%%%%%%%%%%%%%%%%%%%%%%%%%%%%%%%%%%%%%%%%%%%%%%%%%

\end{document}